\documentclass{article}

\usepackage{microtype}
\usepackage{graphicx}
\usepackage{booktabs}
\usepackage{multirow}
\usepackage{hyperref}

\usepackage[preprint]{icml2026}

\usepackage{amsmath}
\usepackage{amssymb}

\newcommand{\pass}{\mathrm{pass}@1}

\icmltitlerunning{Blind Resampling Outperforms Self-Repair in Small Code Models}

\begin{document}

\twocolumn[
\icmltitle{Try Again, Don't Look Back:\\
Blind Resampling Outperforms Self-Repair in Small Code Models}

\begin{icmlauthorlist}
\icmlauthor{Yuvraj Verma}{ind}
\end{icmlauthorlist}

\icmlaffiliation{ind}{Independent Researcher, India}
\icmlcorrespondingauthor{Yuvraj Verma}{yuvrajverma282004@gmail.com}

\icmlkeywords{code generation, self-repair, large language models, ablation
study, anchoring}

\vskip 0.3in
]

\printAffiliationsAndNotice{%
  ORCID: \href{https://orcid.org/0009-0004-2138-3159}{0009-0004-2138-3159}}


\begin{abstract}
Self-repair---returning a failed program to the model together with its test
output and asking for a correction---is a standard component of code agents,
and is almost always evaluated against a baseline that does not retry at all.
We argue that this comparison confounds the value of the feedback with the
value of the extra attempt. Using a placebo-controlled design on MBPP+ at three
model scales (1.5B, 3B, 7B), we compare four matched-budget retry conditions:
blind resampling, a content-free failure notice, genuine execution feedback,
and feedback augmented with verbal self-reflection. Blind resampling is the
strongest condition below 7B, and remains statistically tied with the best
condition at 7B, while consuming $2.5$--$5.5\times$ fewer tokens; conditioning
on the model's own failed attempt costs $6.1$ points at 1.5B ($p{=}0.006$),
and the informational content of execution feedback adds nothing measurable
over the placebo. We attribute this to \emph{anchoring}:
when shown its previous attempt, a model reproduces a near-identical program in
$33$--$68\%$ of retries, against $2$--$14\%$ under blind resampling. Two further
experiments delimit the effect. Retrieved solutions to \emph{other} tasks change
nothing (bounded to $\pm3.5$ points), which localizes the harm to
self-conditioning rather than context length; and reflection, the only
condition that measurably weakens the anchor, remains dominated on cost.
Replication rules out two competing explanations: the penalty is unchanged at
full precision, and it reproduces on an independent model family. Across six
configurations spanning two families and two precisions, its magnitude is
predicted by baseline quality alone ($r{=}0.96$)---the cost of anchoring is
the cost of committing to a bad first attempt.
\end{abstract}

\section{Introduction}
\label{sec:intro}

A code agent that fails a test has an obvious next move: look at what went
wrong and fix it. Systems built on this idea are widespread
\cite{shinn2023reflexion,madaan2023selfrefine,yang2024sweagent}, and the
evidence offered for them is typically a comparison against single-shot
generation. That comparison cannot separate two mechanisms. Retrying at all
raises the success rate under stochastic decoding, independently of whether the
retry is informed; a loop that consumes $k$ additional samples should be
credited only for what it achieves \emph{beyond} $k$ uninformed samples. Absent
that control, an apparent benefit of feedback may be nothing more than the
benefit of persistence.

We therefore treat self-repair as a treatment requiring a placebo. Our control
is \emph{blind resampling}: the original prompt, re-sampled, with no reference
to the failed attempt. Against this control the standard result inverts. In the
1.5B and 3B settings, showing a model its own failed program makes it
\emph{less} likely to succeed than simply asking again; by 7B the penalty has
closed to a statistical tie. Across the whole range, the execution feedback
that practitioners take to be the active ingredient contributes nothing
detectable beyond a content-free notice that the attempt was wrong.

The mechanism we identify is anchoring. A prompt containing a failed program
biases generation toward local edits of that program, whereas a fresh draw is
free to reach a different region of the solution space. We measure this
directly as the textual similarity between consecutive attempts, and find it
predicts not only the aggregate penalty but the relative ordering of every
condition that conditions on the model's own output---an ordering the measure
was not fitted to reproduce.

\paragraph{Contributions.}
We (i) give a placebo-controlled decomposition of the self-repair feedback
packet across three model scales, using the blind-resampling control that prior
evaluations omit; (ii) show self-repair is dominated by retrying on both
accuracy and token cost below 7B; (iii) identify and measure anchoring as the
mechanism, and show it explains the scale trend; (iv) establish, through a null
result on retrieved cross-task experience, that the penalty is specific to
self-conditioning rather than a generic consequence of longer prompts; (v)
test whether verbal reflection escapes the mechanism, finding that it weakens
the anchor without paying for itself; (vi) rule out quantization and model
family as explanations by replication, and show that the penalty's magnitude is
governed by baseline capability across two families; and (vii) release a
harness in which every experiment was pre-registered and runs on a single
consumer GPU.

\section{Related Work}
\label{sec:related}

\paragraph{Iterative self-correction.}
Reflexion \cite{shinn2023reflexion} and Self-Refine
\cite{madaan2023selfrefine} established verbal self-feedback as a general
technique, and both report gains over non-iterative baselines.
\citet{olausson2023selfrepair} question how much of this survives careful
budget accounting, observing that self-repair is often no better than drawing
additional independent samples. Our design differs in isolating \emph{which
component} of the feedback packet carries the effect, and in doing so across
scale: we separate the act of re-reading one's own program from the
information in the test output, and compare both against an equal-budget blind
control. A parallel line of placebo-controlled work on sub-1.5B models reports
that feedback content beats generic placebos; we find that particular contrast
indistinguishable from zero once blind resampling is in the comparison set.

\paragraph{Execution feedback in agents.}
Agent-computer interfaces \cite{yang2024sweagent} and repository-level
benchmarks \cite{jimenez2024swebench} have made execution central to modern
coding agents, though Agentless \cite{xia2024agentless} shows a fixed
localize--repair--validate pipeline is competitive without open-ended agency.
Recent work at frontier scale reports that prohibiting execution entirely costs
little while saving substantial token budget. Our 7B results are consistent
with that regime; our smaller scales show something stronger, namely that the
feedback is actively harmful.

\paragraph{Retrieval and accumulated experience.}
Retrieval-augmented prompting \cite{liu2024rag4code} and experience-driven
issue resolution \cite{zhang2024codagentexp} report benefits from relevant
exemplars, and self-improving systems such as the Darwin G\"odel Machine
\cite{zhang2025dgm} motivate the broader agenda of agents that learn from their
own histories. We find no effect from lexically retrieved exemplars in
small-model code synthesis, and use that null to constrain the interpretation
of our main result.

\paragraph{Evaluation methodology.}
We report $\pass$ \cite{chen2021codex} on EvalPlus \cite{liu2023evalplus},
which augments HumanEval \cite{chen2021codex} and MBPP
\cite{austin2021mbpp} with substantially more tests. Inference follows
McNemar's exact test \cite{mcnemar1947}, bootstrap confidence intervals
\cite{efron1979bootstrap}, Wilson intervals \cite{wilson1927}, and Holm
correction \cite{holm1979}; retrieval uses BM25 \cite{robertson2009bm25}.

\section{Method}
\label{sec:method}

\paragraph{Models and data.}
We evaluate Qwen2.5-Coder \cite{hui2024qwencoder} at 1.5B, 3B and 7B
(Instruct, $Q4\_K\_M$), served locally on a single 6\,GB consumer GPU. MBPP+
(378 tasks) is our primary benchmark; HumanEval+ (164) is reported as
secondary and is underpowered for our contrasts, leaving only 32--75 failing
tasks per cell against 110--164 on MBPP+. Because EvalPlus supplies inputs and
a reference implementation rather than stored outputs, we score by differential
testing: a candidate passes an input when it agrees with the reference on that
input.

\paragraph{Conditions.}
The baseline is greedy single-shot generation. When a program fails, a repair
attempt appends one of four suffixes to the original instruction; the
conditions are identical in every other respect.

\begin{itemize}\itemsep2pt
  \item \textbf{resample} --- nothing is appended. Measures the value of an
  additional draw.
  \item \textbf{placebo} --- the failed program and a content-free notice that
  it was incorrect. Measures the value of re-reading one's own attempt.
  \item \textbf{feedback} --- the failed program and the actual test output
  (failing input, expected value, observed value). Measures the value of the
  diagnostic content.
  \item \textbf{reflect} --- as feedback, preceded by an instruction to
  diagnose the failure in prose before rewriting. Measures the value of
  explicit verbal reasoning.
\end{itemize}

Reading these in sequence decomposes the feedback packet: the step from
\textsf{resample} to \textsf{placebo} isolates self-exposure, and the step from
\textsf{placebo} to \textsf{feedback} isolates diagnostic information.

\paragraph{Decoding.}
Retries sample at temperature $0.8$, applied identically in all conditions;
only the shared $k{=}0$ point is greedy. This is a requirement rather than a
preference. Under greedy decoding the model returns a byte-identical program on
every iteration, which would reduce \textsf{resample} to a no-op and flatten
all curves by construction---an artifact we encountered and corrected before
running the experiment.

\paragraph{Nested evaluation.}
A single trajectory of $k_{\max}{=}8$ records a verdict after every iteration,
so $\pass$ at all $k \le 8$ is read from one run:
\begin{equation}
\pass(k) \;=\; \frac{1}{N}\sum_{i=1}^{N}
  \mathbb{1}\!\left[\exists\, j \le k :\ a_{ij} \text{ passes}\right],
\end{equation}
where $a_{ij}$ is the $j$-th attempt on task $i$. Besides being cheaper than
independent runs per $k$, this yields tighter pairing: the $k{=}1$ and $k{=}8$
arms share an identical trajectory prefix rather than merely a task set. All
repair arms begin from the same baseline run, making the pairing between
baseline and arm exact, and a task that succeeds consumes no further attempts.

\paragraph{Inference.}
Each paired contrast is tested with McNemar's exact test on discordant pairs
\cite{mcnemar1947} and accompanied by a task-level bootstrap interval on the
difference \cite{efron1979bootstrap}. We apply Holm correction
\cite{holm1979} across the primary contrasts at $\alpha{=}0.05$ and report
Wilson intervals \cite{wilson1927} on pass rates. Hypotheses, decision rules
and falsification criteria were registered before each run.

\section{Retrying Beats Repairing}
\label{sec:repair}

\begin{figure*}[t]
  \centering
  \includegraphics[width=\textwidth]{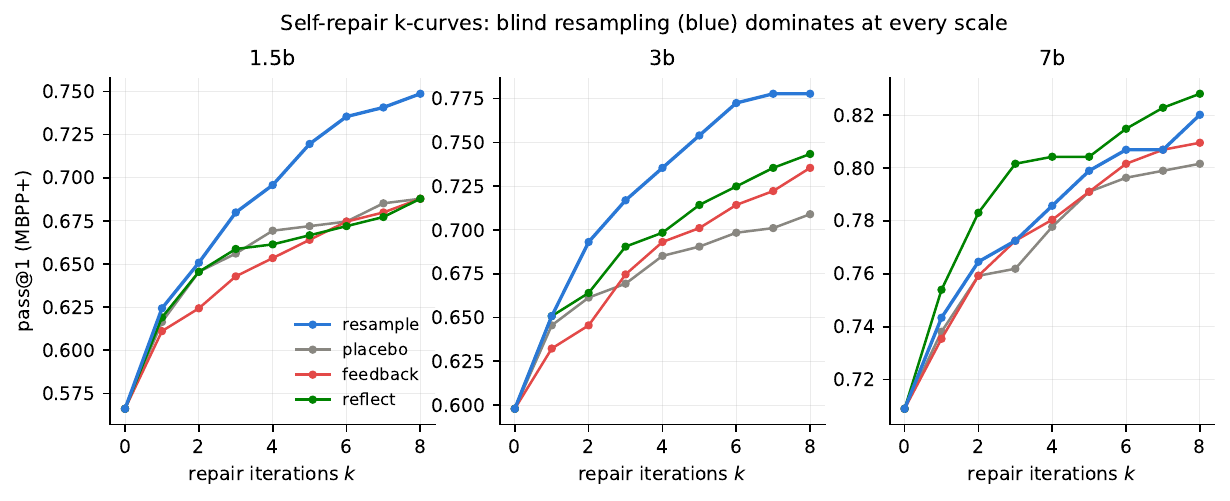}
  \caption{$\pass$ on MBPP+ against repair iterations $k$. Blind resampling
  (blue) dominates the conditions that expose the model to its own failed
  program at 1.5B and 3B; at 7B the conditions converge and reflection edges
  ahead. Each curve is nested from a single $k{=}8$ trajectory.}
  \label{fig:kcurves}
\end{figure*}

Figure~\ref{fig:kcurves} shows the iteration curves and Table~\ref{tab:main}
the endpoint at $k{=}8$. Blind resampling attains the highest pass rate at 1.5B
and 3B and is statistically indistinguishable from the best condition at 7B.
Both pre-registered hypotheses concerning feedback are refuted.

\begin{figure}[t]
  \centering
  \includegraphics[width=\columnwidth]{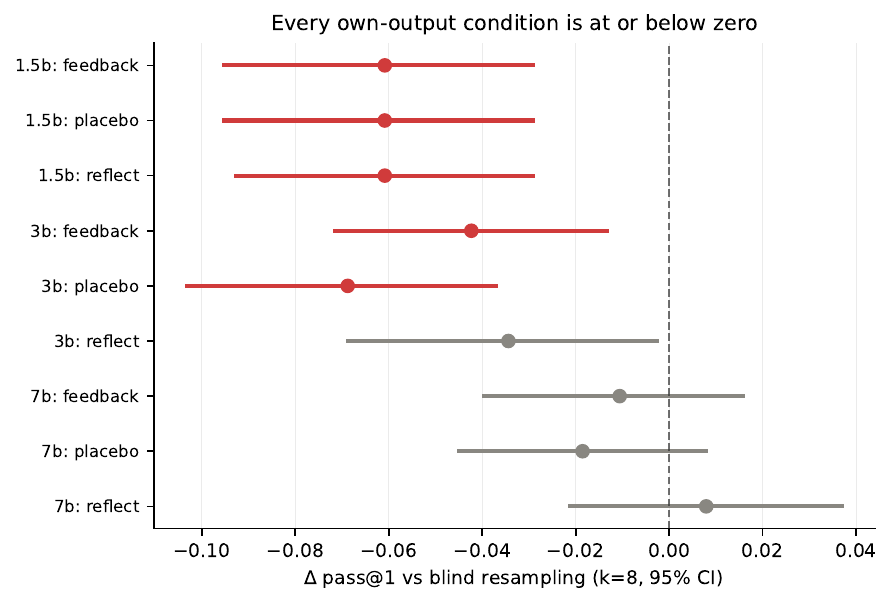}
  \caption{Paired differences in $\pass$ against blind resampling at $k{=}8$,
  with $95\%$ bootstrap intervals. Red marks a significant loss after Holm
  correction, grey a non-significant difference.}
  \label{fig:forest}
\end{figure}

The informational content of execution feedback is not detectable: the contrast
\textsf{feedback} $-$ \textsf{placebo} is $+0.000$, $+0.026$ and $+0.008$ across
scale, none significant. Exposure to one's own failed program is harmful rather
than neutral: \textsf{placebo} $-$ \textsf{resample} is $-0.061$ ($p{=}0.006$)
at 1.5B and $-0.069$ ($p{<}0.001$) at 3B. Figure~\ref{fig:forest} summarizes
the nine contrasts; none favors an own-output condition at any scale.

\begin{table}[t]
  \centering
  \small
  \caption{$\pass$ on MBPP+ at $k{=}8$ with Wilson intervals, alongside
  output-token cost. Best per model in \textbf{bold}. The strongest condition
  is also the cheapest at every scale.}
  \label{tab:main}
  \begin{tabular}{llcc}
    \toprule
    Model & Condition & $\pass$ & Tokens \\
    \midrule
    \multirow{4}{*}{1.5B}
      & resample  & \textbf{0.749} & \textbf{63k} \\
      & placebo   & 0.688 & 90k \\
      & feedback  & 0.688 & 129k \\
      & reflect   & 0.688 & 345k \\
    \midrule
    \multirow{4}{*}{3B}
      & resample  & \textbf{0.778} & \textbf{76k} \\
      & placebo   & 0.709 & 89k \\
      & feedback  & 0.735 & 84k \\
      & reflect   & 0.743 & 284k \\
    \midrule
    \multirow{4}{*}{7B}
      & resample  & 0.820 & \textbf{49k} \\
      & placebo   & 0.802 & 59k \\
      & feedback  & 0.810 & 55k \\
      & reflect   & \textbf{0.828} & 122k \\
    \bottomrule
  \end{tabular}
\end{table}

\begin{figure}[t]
  \centering
  \includegraphics[width=0.95\columnwidth]{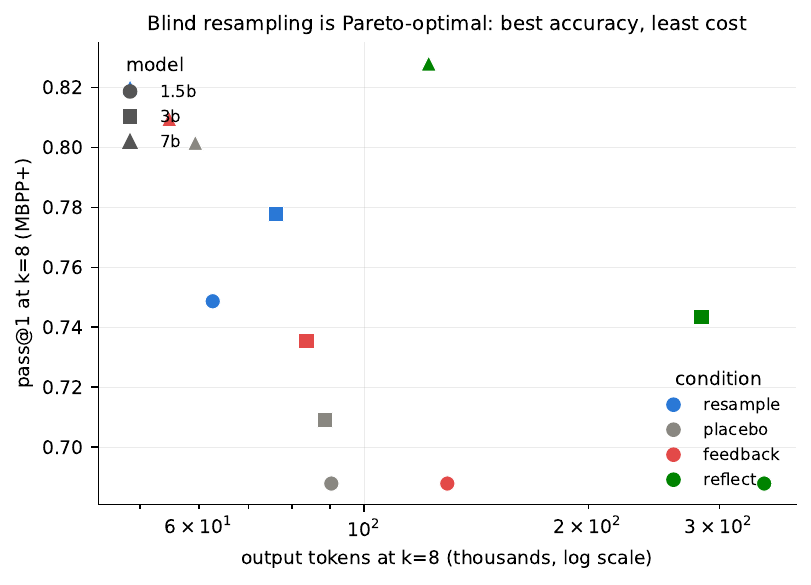}
  \caption{Pass rate at $k{=}8$ against output tokens (log scale). Blind
  resampling is Pareto-optimal at all three scales.}
  \label{fig:pareto}
\end{figure}

\paragraph{Cost.}
Blind resampling is the cheapest condition at every scale
(Figure~\ref{fig:pareto}), partly because succeeding more often lets it exit
earlier. Below 7B this makes the comparison one-sided: at 1.5B and 3B the
own-output conditions are strictly Pareto-dominated, being worse on accuracy
\emph{and} more expensive. At 7B the picture is a genuine trade-off rather
than a dominance: reflection buys $+0.8$ points, which is not significant, for
$2.5\times$ the output tokens. Blind resampling remains on the Pareto frontier
at all three scales.

\paragraph{Where the gains are.}
Improvement is concentrated in the first two iterations: $46$--$53\%$ of all
attainable gain is realized by $k{=}2$, and marginal returns approach zero by
$k{=}7$. Budgets beyond roughly two iterations are poorly spent regardless of
condition.

\section{Anchoring}
\label{sec:anchoring}

\begin{figure}[t]
  \centering
  \includegraphics[width=\columnwidth]{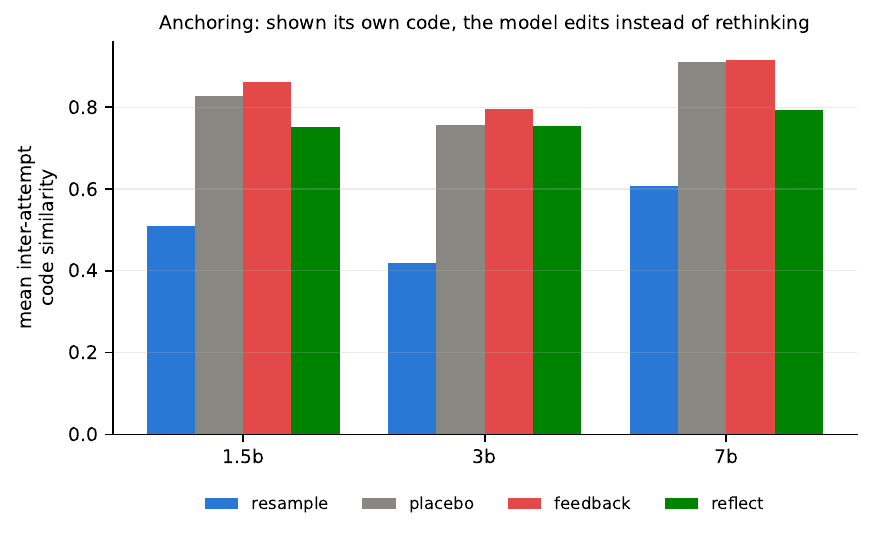}
  \caption{Mean similarity between consecutive attempts. Conditions that expose
  the model to its own program reproduce near-identical solutions far more
  often than blind resampling.}
  \label{fig:anchoring}
\end{figure}

We hypothesized before measuring that conditioning on a failed program biases
the model toward local edits. To test this we compute the mean textual
similarity between consecutive attempts, using a longest-matching-subsequence
ratio over extracted code (Figure~\ref{fig:anchoring}).

The effect is large and consistent. Similarity rises by $0.30$ to $0.38$
relative to blind resampling at every scale, and the proportion of retries that
are near-identical ($>0.95$) rises from $2$--$14\%$ to $33$--$68\%$. Shown its
own code, the model edits; asked afresh, it reconsiders.

This also accounts for the scale trend, which the original hypothesis did not
anticipate. Anchoring \emph{strength} is roughly constant across scale, but its
\emph{cost} declines: a larger model's failed first attempt is a better object
to anchor upon. The diminishing penalty at 7B therefore reflects anchoring
becoming inexpensive rather than feedback becoming useful.

\begin{figure}[t]
  \centering
  \includegraphics[width=0.9\columnwidth]{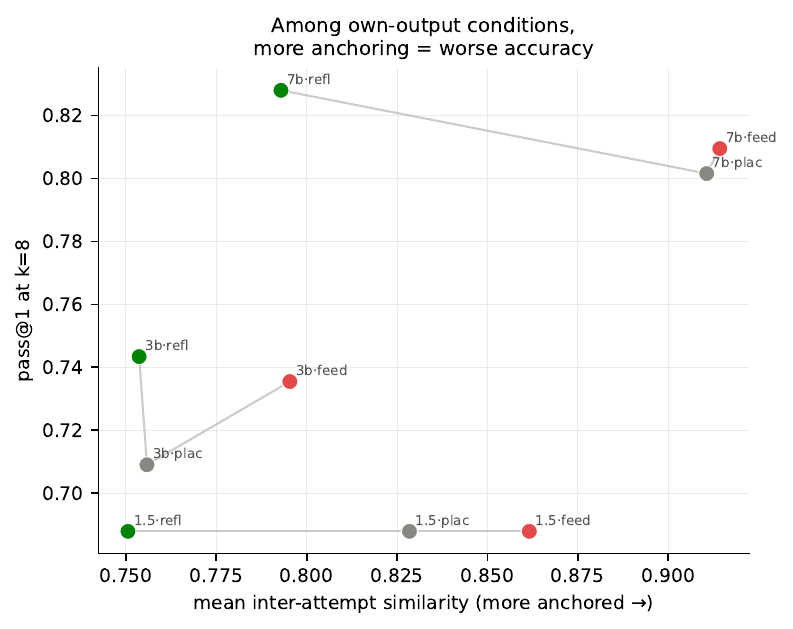}
  \caption{Within each model, greater inter-attempt similarity corresponds to
  lower $\pass$. Lines connect the three own-output conditions.}
  \label{fig:ordering}
\end{figure}

\paragraph{A partial ordering the measure was not fitted to.}
Among the three conditions that expose the model to its own output, the
\emph{least} anchored condition is the best performer at all three scales
(Figure~\ref{fig:ordering}): reflection reduces similarity by $0.04$--$0.12$
relative to feedback and is correspondingly the strongest of the three. The
measure does not, however, resolve the remaining pair. Placebo and feedback
differ in anchoring by only $0.034$, $0.039$ and $0.003$ across scale---within
the noise of the metric---and their accuracy ordering does not follow, with
placebo the weaker of the two at 3B and 7B despite being marginally less
anchored. At 1.5B the three conditions are exactly tied in accuracy, so no
ordering exists to recover. We therefore read this as directional support for
the mechanism at the resolved end of the range rather than as a rank
correspondence, and we do not claim the measure explains differences between
conditions whose anchoring is indistinguishable.

\section{Cross-Task Experience Has No Effect}
\label{sec:memory}

A natural objection is that long prompts simply distract small models, in which
case adding any code to the context should hurt. We test this by prepending two
solved exemplars---drawn either at random or by BM25 relevance from the model's
own past successes---under leave-one-out, so a task never retrieves itself. An
audit conducted before the experiment found that only $1.2$--$1.7\%$ of
retrieved exemplars exceed $0.8$ similarity to the reference solution for the
query task.

\begin{table}[t]
  \centering
  \small
  \caption{$\pass$ on MBPP+ with retrieved experience. All nine paired
  contrasts are non-significant after Holm correction, bounding any effect to
  $\pm3.5$ points.}
  \label{tab:memory}
  \begin{tabular}{lccc}
    \toprule
    Model & none & random & retrieved \\
    \midrule
    1.5B & 0.566 & 0.542 & 0.542 \\
    3B   & 0.598 & 0.595 & 0.606 \\
    7B   & 0.709 & 0.728 & 0.696 \\
    \bottomrule
  \end{tabular}
\end{table}

The outcome is a clean null (Table~\ref{tab:memory}). Read alongside
Section~\ref{sec:repair} it is informative: both interventions add a comparable
quantity of code to the context, yet a model's own failed attempt costs $6.1$
points while other tasks' successful solutions cost nothing. The harm is
therefore specific to self-conditioning, which is what anchoring predicts and
what a context-dilution account does not. We note that this null concerns
lexically retrieved exemplars in code synthesis, where the required output
format is already specified in the prompt; it does not contradict retrieval
gains reported on classification tasks, where exemplars also teach a
convention.

\section{Reflection Weakens the Anchor}
\label{sec:reflect}

Verbal reasoning \cite{shinn2023reflexion} is the natural candidate for
escaping the mechanism, since articulating \emph{why} an approach failed might
license abandoning it. It partially does. Reflection is the only own-output
condition that measurably reduces anchoring---similarity falls by $0.11$,
$0.04$ and $0.12$ relative to feedback, and near-identical regeneration drops
by roughly a third---and it is correspondingly the strongest of the three
(Table~\ref{tab:main}).

It nevertheless does not pay for itself. Reflection loses to blind resampling
by $6.1$ points at 1.5B ($p{=}0.006$), loses insignificantly at 3B, and merely
matches it at 7B while consuming $2.5\times$ the output tokens ($5.5\times$ at
1.5B). It is best understood as a partial and expensive mitigation of a problem
that is avoided entirely by not conditioning on one's own output.

\section{Replication: Precision and Model Family}
\label{sec:replication}

Two explanations compete with the anchoring account. The penalty might be an
artifact of $Q4$ quantization, which is documented to degrade in-context
learning more than unconditional generation---and the conditions we find
wanting are exactly those that place extra material in context. Or it might be
specific to the Qwen pre-training distribution. We test both by re-running the
decomposition with one factor changed at a time: unquantized weights at 1.5B,
and an independent model family (DeepSeek-Coder) at 1.3B and 6.7B. Neither
explanation survives, and the family replication yields a sharper account of
the scale trend.

\paragraph{Precision.}
Repeating the experiment with unquantized (FP16) weights, which also removes
CPU offload because the 3.1\,GB model fits the GPU entirely, reproduces every
quantity (Table~\ref{tab:replication}). The headline contrast is
$-0.063$ at FP16 against $-0.061$ at $Q4$, still significant
($p{=}0.0008$); \textsf{feedback} $-$ \textsf{placebo} remains null
($+0.016$, n.s.); and the anchoring measure is unchanged to within $0.003$,
with placebo identical at $0.828$ in both. The objection predicts the penalty
should shrink toward zero without quantization. It does not shrink at all, and
the placebo penalty is in fact slightly larger ($-0.079$). We therefore regard
precision as excluded.

\begin{table}[t]
  \centering
  \small
  \caption{Replication at 1.5B on MBPP+ ($k{=}8$). Changing precision leaves
  every quantity intact, including the anchoring measure.}
  \label{tab:replication}
  \begin{tabular}{lccc}
    \toprule
    Quantity & $Q4$ & FP16 & $\Delta$ \\
    \midrule
    resample                 & 0.749 & 0.746 & $-0.003$ \\
    placebo                  & 0.688 & 0.667 & $-0.021$ \\
    feedback                 & 0.688 & 0.683 & $-0.005$ \\
    \textbf{feedback $-$ resample} & $\mathbf{-0.061}$ & $\mathbf{-0.063}$ & $-0.002$ \\
    \midrule
    anchoring, placebo       & 0.828 & 0.828 & $0.000$ \\
    anchoring, feedback      & 0.862 & 0.865 & $+0.003$ \\
    \bottomrule
  \end{tabular}
\end{table}

\paragraph{Model family.}
Repeating the decomposition on DeepSeek-Coder---a different laboratory,
pre-training corpus and tokenizer---reproduces the penalty at both scales
tested. At 1.3B, \textsf{placebo} $-$ \textsf{resample} is $-0.138$
($p<10^{-4}$) and \textsf{feedback} $-$ \textsf{resample} is $-0.114$
($p<10^{-4}$); at 6.7B the corresponding values are $-0.045$ ($p{=}0.007$)
and $-0.029$ (n.s.). As with Qwen, \textsf{feedback} $-$ \textsf{placebo} is
null in both cases. The effect is therefore not a property of one
pre-training distribution.

\begin{figure}[t]
  \centering
  \includegraphics[width=0.92\columnwidth]{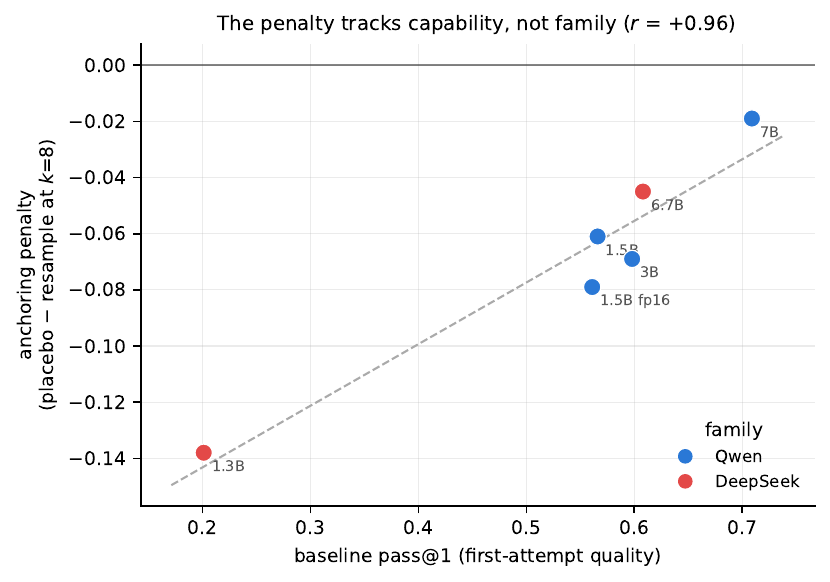}
  \caption{Anchoring penalty against baseline quality. Points from the two
  model families interleave along a single trend, indicating that capability
  rather than family governs the penalty.}
  \label{fig:capability}
\end{figure}

\paragraph{The penalty tracks capability, not scale or family.}
Pooling all six configurations (Figure~\ref{fig:capability} and
Table~\ref{tab:capability}), the anchoring
penalty is strongly predicted by baseline quality alone
($r{=}{+}0.96$ against \textsf{placebo} $-$ \textsf{resample};
$r{=}{+}0.95$ against \textsf{feedback} $-$ \textsf{resample}). Two
comparisons separate the candidate explanations. Qwen-3B and DeepSeek-6.7B
have nearly identical baselines ($0.598$ and $0.608$) and nearly identical
penalties ($-0.069$ and $-0.045$) despite different families, so family
explains little. DeepSeek-1.3B and DeepSeek-6.7B share a family yet differ
$3.1\times$ in penalty ($-0.138$ against $-0.045$), so capability explains
much. This is what the mechanism predicts: the cost of committing to a
previous attempt is the cost of committing to a \emph{bad} attempt, and so
scales with how bad that attempt typically is. The apparent scale trend of
Section~\ref{sec:repair} is thus a capability effect, now observed across two
independent model families. We report the correlation as descriptive support
rather than a fitted law: six unevenly spaced points do not warrant more.

\begin{table}[t]
  \centering
  \small
  \caption{Anchoring penalty against baseline quality across two families and
  two precisions. Ordered by baseline; $r{=}{+}0.96$.}
  \label{tab:capability}
  \begin{tabular}{llcc}
    \toprule
    Model & Family & $\pass$ & placebo $-$ resample \\
    \midrule
    1.3B      & DeepSeek & 0.201 & $-0.138$ \\
    1.5B FP16 & Qwen     & 0.561 & $-0.079$ \\
    1.5B $Q4$ & Qwen     & 0.566 & $-0.061$ \\
    3B        & Qwen     & 0.598 & $-0.069$ \\
    6.7B      & DeepSeek & 0.608 & $-0.045$ \\
    7B        & Qwen     & 0.709 & $-0.019$ \\
    \bottomrule
  \end{tabular}
\end{table}

\section{Discussion}
\label{sec:discussion}

\paragraph{Instrumentation.}
Four measurement faults were identified and corrected before they could reach a
result, each of which would have produced a plausible but incorrect number:
code extraction that stripped leading indentation; scored run manifests emitted
for an unavailable model; a token budget that bound one arm and not another, so
that a raw pass-rate gap measured budget fit rather than capability; and greedy
retries returning identical programs. We verified that no arm was
budget-limited, with zero budget exhaustion across all repair and memory
responses.

\paragraph{Threats to validity.}
Arms differ in token consumption because early exit on success is intended and
cost is treated as an outcome; the direction of that difference favors the
losing arms, since the winning condition spends least. Feedback quotes failing
inputs drawn from the same suite used for scoring, so any leakage would inflate
the condition that lost.

Our scope is limited to one benchmark family and function-level synthesis,
with lexical rather than semantic retrieval and a single reflection prompt
formulation. Two further concerns---that the effect is an artifact of
quantization, or specific to one model family---are addressed directly in
Section~\ref{sec:replication}.

\paragraph{The control determines the conclusion.}
For any capability that adds context to a prompt, the appropriate control is an
equally expensive but uninformative alternative rather than the absence of the
capability. Measured against a no-repair baseline our feedback condition appears
to gain $12.2$ points at 1.5B; measured against blind resampling at matched
budget it loses. The treatment is unchanged---only the comparison moved.

\section{Conclusion}
\label{sec:conclusion}

At 1.5B and 3B, self-repair on function-level code tasks is worse than blind
resampling on both accuracy and cost, because conditioning on a failed attempt
anchors the model to an unsuccessful approach; by 7B the penalty has closed to
a tie, with reflection buying a non-significant $+0.8$ points for
$2.5\times$ the tokens. Seven-billion parameters thus reads as a transitional
point at which anchoring becomes cheap rather than one at which feedback
becomes informative---the feedback-versus-placebo contrast remains null
throughout. Replication shows the effect is neither a quantization artifact nor
a property of one model family, and that its magnitude is set by baseline
capability: the penalty is the cost of committing to a bad first attempt, so it
fades as first attempts improve. Retrieved cross-task experience produces no
effect, which localizes the mechanism to self-conditioning. A practical reading
for small-model agents:
retry rather than patch, cap iteration budgets near $k{\approx}2$, and treat
execution feedback as a costed resource rather than a default.

The capability relationship suggests the sharpest remaining test: a model whose
first attempts are strong enough that anchoring should cost nothing, which our
correlation places not far beyond the 7B point. Substituting semantic retrieval
or distilled lessons for raw exemplars would likewise test whether cross-task
experience remains inert under a stronger retriever.

\section*{Impact Statement}

This paper advances the empirical understanding of when iterative
self-correction helps code-generating models. Its principal practical
consequence is a reduction in compute: we find that the cheapest strategy is
also the most accurate at the scales studied, so practitioners who adopt the
recommendation will spend fewer tokens for equal or better results, with a
corresponding reduction in energy use. A methodological consequence is that
capabilities added to agent prompts should be evaluated against
equally-expensive uninformative controls; adopting this standard should reduce
the number of overstated capability claims in the literature. We see no
specific ethical risk arising from this work beyond those already general to
automated code generation, and our findings, being cautionary about an existing
technique, do not enable new harmful capability.

\section*{Acknowledgements}

The author declares no competing financial interests and received no funding
for this work. All experiments were run on personal hardware.

An AI coding assistant was used during this project to help implement the
experimental harness and to draft portions of the manuscript. All experimental
designs, hypotheses and decision rules were pre-registered before the
corresponding runs; every reported number derives from a recorded run manifest
and a replayable execution trace, and the author has verified the results and
takes full responsibility for the content of this paper.

\section*{Code and Data Availability}

The harness, pre-registrations, run traces and figure-generation scripts are
available at
\url{https://github.com/vermayuvraj/self-improving-agent}.
Every figure and table is regenerated from recorded run artifacts.

\bibliography{refs}

@article{chen2021codex,
  title   = {Evaluating Large Language Models Trained on Code},
  author  = {Chen, Mark and Tworek, Jerry and Jun, Heewoo and others},
  journal = {arXiv preprint arXiv:2107.03374},
  year    = {2021}
}

@inproceedings{austin2021mbpp,
  title     = {Program Synthesis with Large Language Models},
  author    = {Austin, Jacob and Odena, Augustus and Nye, Maxwell and others},
  booktitle = {arXiv preprint arXiv:2108.07732},
  year      = {2021}
}

@inproceedings{liu2023evalplus,
  title     = {Is Your Code Generated by {ChatGPT} Really Correct? Rigorous
               Evaluation of Large Language Models for Code Generation},
  author    = {Liu, Jiawei and Xia, Chunqiu Steven and Wang, Yuyao and Zhang, Lingming},
  booktitle = {Advances in Neural Information Processing Systems (NeurIPS)},
  year      = {2023}
}

@article{hui2024qwencoder,
  title   = {{Qwen2.5-Coder} Technical Report},
  author  = {Hui, Binyuan and Yang, Jian and Cui, Zeyu and others},
  journal = {arXiv preprint arXiv:2409.12186},
  year    = {2024}
}

@inproceedings{shinn2023reflexion,
  title     = {Reflexion: Language Agents with Verbal Reinforcement Learning},
  author    = {Shinn, Noah and Cassano, Federico and Berman, Edward and
               Gopinath, Ashwin and Narasimhan, Karthik and Yao, Shunyu},
  booktitle = {Advances in Neural Information Processing Systems (NeurIPS)},
  year      = {2023}
}

@article{madaan2023selfrefine,
  title   = {Self-Refine: Iterative Refinement with Self-Feedback},
  author  = {Madaan, Aman and Tandon, Niket and Gupta, Prakhar and others},
  journal = {Advances in Neural Information Processing Systems (NeurIPS)},
  year    = {2023}
}

@article{olausson2023selfrepair,
  title   = {Is Self-Repair a Silver Bullet for Code Generation?},
  author  = {Olausson, Theo X. and Inala, Jeevana Priya and Wang, Chenglong and
             Gao, Jianfeng and Solar-Lezama, Armando},
  journal = {International Conference on Learning Representations (ICLR)},
  year    = {2024}
}

@article{yang2024sweagent,
  title   = {{SWE-agent}: Agent-Computer Interfaces Enable Automated Software
             Engineering},
  author  = {Yang, John and Jimenez, Carlos E. and Wettig, Alexander and others},
  journal = {Advances in Neural Information Processing Systems (NeurIPS)},
  year    = {2024}
}

@article{jimenez2024swebench,
  title   = {{SWE-bench}: Can Language Models Resolve Real-World {GitHub} Issues?},
  author  = {Jimenez, Carlos E. and Yang, John and Wettig, Alexander and others},
  journal = {International Conference on Learning Representations (ICLR)},
  year    = {2024}
}

@article{xia2024agentless,
  title   = {Agentless: Demystifying {LLM}-based Software Engineering Agents},
  author  = {Xia, Chunqiu Steven and Deng, Yinlin and Dunn, Soren and Zhang, Lingming},
  journal = {arXiv preprint arXiv:2407.01489},
  year    = {2024}
}

@article{zhang2024codagentexp,
  title   = {{SWE-Exp}: Experience-Driven Software Issue Resolution},
  author  = {Zhang, Silin and others},
  journal = {arXiv preprint arXiv:2507.23361},
  year    = {2025}
}

@article{zhang2025dgm,
  title   = {{Darwin} {G{\"o}del} Machine: Open-Ended Evolution of
             Self-Improving Agents},
  author  = {Zhang, Jenny and Hu, Shengran and Lu, Cong and Lange, Robert and
             Clune, Jeff},
  journal = {arXiv preprint arXiv:2505.22954},
  year    = {2025}
}

@article{liu2024rag4code,
  title   = {Retrieval-Augmented Generation for Code Generation: A Survey},
  author  = {Liu, others},
  journal = {arXiv preprint},
  year    = {2024}
}

@article{mcnemar1947,
  title   = {Note on the Sampling Error of the Difference Between Correlated
             Proportions or Percentages},
  author  = {McNemar, Quinn},
  journal = {Psychometrika},
  volume  = {12},
  number  = {2},
  pages   = {153--157},
  year    = {1947}
}

@article{holm1979,
  title   = {A Simple Sequentially Rejective Multiple Test Procedure},
  author  = {Holm, Sture},
  journal = {Scandinavian Journal of Statistics},
  volume  = {6},
  number  = {2},
  pages   = {65--70},
  year    = {1979}
}

@article{efron1979bootstrap,
  title   = {Bootstrap Methods: Another Look at the Jackknife},
  author  = {Efron, Bradley},
  journal = {The Annals of Statistics},
  volume  = {7},
  number  = {1},
  pages   = {1--26},
  year    = {1979}
}

@article{robertson2009bm25,
  title   = {The Probabilistic Relevance Framework: {BM25} and Beyond},
  author  = {Robertson, Stephen and Zaragoza, Hugo},
  journal = {Foundations and Trends in Information Retrieval},
  volume  = {3},
  number  = {4},
  pages   = {333--389},
  year    = {2009}
}

@article{wilson1927,
  title   = {Probable Inference, the Law of Succession, and Statistical
             Inference},
  author  = {Wilson, Edwin B.},
  journal = {Journal of the American Statistical Association},
  volume  = {22},
  number  = {158},
  pages   = {209--212},
  year    = {1927}
}
\bibliographystyle{icml2026}

\end{document}